\documentclass[twocolumn,aps,superscriptaddress,showpacs]{revtex4-2}

\usepackage{amssymb}
\usepackage{amsmath}
\usepackage{mathrsfs}
\usepackage{graphicx}
\usepackage{subfigure}
\usepackage[normalem]{ulem}
\usepackage[dvips]{color}

\begin{document}

\title{Axion effects on quark matter and quark-matter cores in massive hybrid stars}
\author{He Liu}
\email{liuhe@qut.edu.cn}
\affiliation{Science School, Qingdao University of Technology, Qingdao 266000, China}
\author{Yu-Heng Liu} 
\affiliation{Science School, Qingdao University of Technology, Qingdao 266000, China}
\author{Yong-Hang Yang}
\affiliation{Science School, Qingdao University of Technology, Qingdao 266000, China}
\author{Min Ju}
\email{jumin@upc.edu.cn}
\affiliation{School of Science, China University of Petroleum (East China), Qingdao 266580, China}
\author{Xu-Hao Wu}
\email{wuhaoysu@ysu.edu.cn}
\affiliation{School of Science, Yanshan University, Qinhuangdao 066004, China}
\author{Hong-Ming Liu}
\email{liuhongming13@126.com}
\affiliation{Science School, Qingdao University of Technology, Qingdao 266000, China}
\author{Peng-Cheng Chu}
\email{kyois@126.com}
\affiliation{Science School, Qingdao University of Technology, Qingdao 266000, China}
\date{\today}
\begin{abstract}
Using a three-flavor Nambu--Jona-Lasinio model to describe the charge-parity violating effects through axion field, we investigate the axion effects on quark matter and quark-matter cores in massive hybrid stars.  Within the range from 0 to $\pi $, the axion field $a/f_a$ decreases the baryon chemical potential of the first-order phase transition, leading to an increase in normalized pressure and stiffening of the quark matter equation of state. The effect of axions on hybrid star matter that includes the hadron-quark phase transition is contrary to expectations. The axion field shifts the onset of the hadron-quark mixed phase to lower densities but slightly softens the equation of state of the mixed phase matter, which also results in a slight decrease in the maximum mass and corresponding radius of the hybrid stars. However, we also find that the lowering of the onset of the mixed phase significantly increases the radius and mass of the quark-matter core in the hybrid star. Therefore, our results indicate that with axion effects, a sizable quark-matter core can appear in $2M_{\odot}$ massive neutron stars. 
\end{abstract}

\maketitle
\textit{Introduction.} Recent inspiring progress has been made in astrophysical observations on neutron stars (NSs), including the measurements of masses and radii of the millisecond pulsars PSR J0030+0451~\cite{Ril19,Mil19} and PSR J0740+6620~\cite{Cro20,Ril21,Mil21} using the Neutron Star Interior Composition Explorer (NICER), as well as the multi-messenger observations of gravitational-wave events GW170817~\cite{Abb17} and GW190814~\cite{Abb20} by the LIGO/Virgo Collaboration. These observations of massive NSs have advanced our understanding of the equation of state (EOS) for dense, strongly interacting matter. It is generally believed that a first-order phase transition from hadronic matter to quark matter at high baryon densities may occur in the interior of NSs~\cite{Web05,Aok06}. Some recent studies, for instance, have also shown that the quark-matter core can appear in massive NS~\cite{Ann20}, and the presence of a first-order phase transition from hadronic to quark matter can imprint signatures in binary NS merger observations~\cite{Bau19,Hua22}. Binary NS merger events have recently emerged as a new tool for probing beyond-the-standard-model (BSM) particles, such as axions and axionlike particles (ALPs)~\cite{Dev24,Dia24}, CP-even scalars~\cite{Dev22}, and dark photons~\cite{Dia22}. The hot and dense matter in a binary NS merger remnant would efficiently produce ALPs coupling to photons. However, as a candidate for cold dark matter~\cite{Wei78,Wil78,Din83}, axion was originally introduced to explain the violation of combined symmetries of charge conjugation and parity (CP violation) in quantum chromodynamics (QCD)\cite{Wei78,Wil78}, which have also been associated with stellar evolution~\cite{Raf91,Jan96} and the anomalous stellar cooling problem~\cite{Sed16,Sed19}. Hence, this new/exotic matter state would also be produced in the cold NSs before they merge and affect their properties.

\textit{Methods.} Due to the non-Abelian nature of the gauge fields, QCD allows the topologically nontrivial Chern-Simons term $\mathcal{L}_\theta \sim \theta \text{Tr} G_{\mu\nu} \tilde{G}^{\mu\nu}$. This term has important quantum mechanical consequences, as it leads to an explicit breaking of the CP symmetry in QCD for a non-vanishing value $\theta$~\cite{Col79}. The axion as a pseudo-Goldstone boson from spontaneous breaking of the Peccei-Quinn (PQ) symmetry~\cite{Pec77L,Pec77D} is an elegant mechanism to solve the strong CP violating problem. The properties of QCD axion in a hot and dense medium have recently been studied in the framework of the three-flavor Nambu--Jona-Lasinio (NJL) model for quark matter~\cite{Boo09,Cha12,Abh21}.  The CP violating effects through the axion field can be included in the Kobayashi-Maskawa-t' Hooft (KMT) determinant term~\cite{Hoo76}. Thus, the Lagrangian density of the three flavor NJL model with axion fields can be given by
\begin{eqnarray}\label{eq1}
\mathcal{L}&=& \bar{\psi}(i\rlap{\slash}\partial-\hat{m})\psi 
+\frac{G_{S}}{2}\sum_{a=0}^{8}[(\bar{\psi}\lambda_{a}\psi)^{2}+(\bar{\psi}i\gamma_{5}\lambda_{a}\psi)^{2}] 
\notag\\
&-& K\{e^{i\frac{a}{f_a}}\det[\bar{\psi}(1+\gamma_{5})\psi]+e^{-i\frac{a}{f_a}}\det[\bar{\psi}(1-\gamma_{5})\psi]\}
\notag\\
&-&\frac{G_{V}}{2}\sum_{a=0}^{8}[(\bar{\psi}\gamma_{\mu}\lambda_{a}\psi)^{2}+(\bar{\psi}\gamma_{5}\gamma_{\mu}\lambda_{a}\psi)^{2}],
\end{eqnarray}
where $\psi = (u, d, s)^T$ represents the quark field with three flavors, $\hat{m} = diag(m_u,m_d,m_s)$ is the current quark mass matrix, and $\lambda_a$ shows the flavor SU(3) Gell-Mann matrices with $\lambda_0 = \sqrt{2/3}I$. $G_S$ and $G_V$ are, respectively, the scalar and vector coupling constants. The $K$ term in Eq. (1) represents the six-point interaction that breaks the axial symmetry $U(1)_A$~\cite{Hoo76}. The normalized axion field in the term $K$ is denoted as $a(x)=\theta(x) f_a$, where $f_a$ is the decay constant of the axion which also represents the PQ symmetry breaking scale. Astrophysical observations, e.g., cooling rate of the SN1987A supernova and black hole superradiance, put stringent constraint on the PQ symmetry-breaking scale with $10^8 \leq f_a \leq 10^{17}$ GeV~\cite{Raf08,Cha18,Bar20}. In our case, we are dealing with a much smaller energy scale than the axion symmetry-breaking energy. Hence, we can take the axion field $a$ to be in its vacuum expectation value, and the interaction between the axion field and the QCD gauge field can now be expressed as $\mathcal{L}_\theta \sim (a/f_a)\text{Tr} G_{\mu\nu} \tilde{G}^{\mu\nu}$. 

In the mean-field approximation, the thermodynamic potential from the finite-temperature field theory can be expressed as
{\small
\begin{eqnarray}\label{eq2}
\Omega&=&-2N_c\sum_{i}\{\int_0^{\Lambda}\frac{d^3p}{(2\pi)^3}E_i+\int\frac{d^3p}{(2\pi)^3}[T\ln(1+e^{-\beta(E_i-\tilde{\mu}_i)})
\notag\\
&+&T\ln(1+e^{-\beta(E_i+\tilde{\mu}_i)})]\}+\sum_i[G_S(\sigma_i^2+\eta_i^2)-G_V \rho_i^2]
\notag\\
&-&4K[ \cos\frac{a}{f_a}(\sigma_u\sigma_d\sigma_s-\sigma_s\eta_u\eta_d-\sigma_u\eta_d\eta_s-\sigma_d\eta_u\eta_s)
\notag\\
&-& \sin\frac{a}{f_a}(\eta_u\eta_d\eta_s-\sigma_u\sigma_s\eta_d-\sigma_d\sigma_s\eta_u-\sigma_u\sigma_d\eta_s)],
\end{eqnarray}}where the factor $2N_c=6$ represents the spin and color degeneracy of the quark, $\beta = 1/T$ is the inverse of the temperature, and $\tilde\mu_i=\mu_i-2G_V\rho_i$ is the effective chemical potential. Note that the first integral, corresponding to the vacuum contribution, is ultraviolet divergence, necessitating a cutoff \(\Lambda\) to regularize the three-momentum integration~\cite{Fuk04}. In the above,
$E_{i} =\sqrt{M_i^{2}+p^{2}}$ with $M_i =\sqrt{{M_i^s}^2+{M_i^p}^2}$ is the single-particle energy for $i$-flavor quark, where $M_i^s$ and $M_i^p$ are the scalar and pseudoscalar contributions of the constituent mass given by the gap equations
\begin{eqnarray}\label{eq3}
M_i^s& =&m_{i}+2G_S\sigma_i+2K[\cos\frac{a}{f_a}(\sigma_j\sigma_k-\eta_j\eta_k)\notag\\
&+&\sin\frac{a}{f_a}(\sigma_j\eta_k+\eta_j\sigma_k)], 
\\ \label{eq4}
 M_i^p&=&2G_S\eta_{i}+2K[\cos\frac{a}{f_{a}}(\eta_j\sigma_k+\sigma_j\eta_k)\notag\\
&+&\sin\frac {a}{f_a}(\eta_j\eta_k-\sigma_j\sigma_k)]
\end{eqnarray}
with$(i,j,k)=(u,d,s)$. $\sigma_i=\langle \bar{\psi_i} \psi_i \rangle$ and $\eta_i=\langle \bar{\psi_i}i\gamma_5 \psi_i \rangle$ represent scalar and pseudoscalar condensates, respectively. In the case of unbroken isospin symmetry, only nonzero $\sigma_i$ and/or $\eta_i$ can arise. When $a/f_a \neq 
0$, a nonzero $\eta_i$ signals that CP invariance is broken, and thus it can serve as an order parameter for the CP-violating phase~\cite{Boo09}. Those two condensates can be obtained as~\cite{Boo09,Cha12}
{\small
\begin{eqnarray}\label{eq5}
\sigma_i&=&-2N_c \Big[\int_0^\Lambda\frac{d^3p}{(2\pi)^3}\frac{M_i^s}{E_i}-\int\frac{d^3p}{(2\pi)^3} \frac{M_i^s}{E_i}(f_i+\bar{f_i}) \Big], \\
\eta_i&=&2N_c\Big[\int_0^\Lambda\frac{d^3p}{(2\pi)^3}\frac{M_i^p}{E_i}-\int\frac{d^3p}{(2\pi)^3}\frac{M_i^p}{E_i}(f_i+\bar{f_i})\Big],
\end{eqnarray}
}where $f_i$ and $\bar{f}_i$ are respectively the Fermi distribution functions of particle and antiparticle. The pressure and energy density can be derived using the thermodynamic relations in the grand canonical ensemble as $P=-(\Omega-\Omega_0)$ and $\varepsilon=\sum_i\mu_i\rho_i-P$, where $\Omega_0$ is introduced to ensure that the pressure and energy density vanish in the vacuum. In the present study, we employ the parameters $m_u =m_d =3.6$ MeV, $m_s =87$ MeV, $G_S\Lambda^2 = 3.6$, $K\Lambda^5 = 8.9$, and the cutoff value in the momentum integral $\Lambda = 750$ MeV given in Refs.~\cite{Bra13,Lut92}. 

To investigate the effects of axions on quark matter in massive neutron stars, we primarily focus on hybrid stars with the hadron-quark phase transition. Nuclear matter in hybrid stars can be described using an improved isospin- and momentum-dependent interaction (ImMDI) model. In our previous study~\cite{Jxu15,Liu22,Liu231}, the ImMDI model is constructed from fitting the properties of cold symmetric nuclear matter (SNM), which is approximately reproduced by the self-consistent Greens function (SCGF) approach~\cite{Car14,Car18} or chiral effective many-body pertubation theory ($\chi$EMBPT) ~\cite{Wel15,Wel16}. The ImMDI model has been extensively used in intermediate-energy heavy-ion reactions to study the properties of asymmetric nuclear matter. In the ImMDI model, we introduce the parameters $x$, $y$ and $z$ to adjust the slope $L$ of symmetry energy, the momentum dependence of the symmetry potential, and the symmetry energy $E_{sym}$ at saturation density, respectively. In order to better focus on axion effects on the properties of quark matter in hybrid stars, we thus choose a fixed parameter set in ImMDI model\cite{Liu231}, $x=-0.3$, $y=32$ MeV, and $z=0$~, that would allow $2.08M_{\odot}$ neutron stars and still satisfy well nuclear matter constraints at saturation density $\rho_0=0.16fm^{-3}$, i.e., the binding energy $E_0(\rho_0) = -15.9$ MeV, the incompressibility $K_0 = 240$ MeV, the symmetry energy $E_{sym}(\rho_0) = 32.5$ MeV, the slope parameter $L=106$ MeV, the isoscalar effective mass $m^{\star}_s = 0.7m$, and the single-particle potential $U_{0,\infty} = 75$ MeV at infinitely large nucleon momentum.

\begin{figure*}[tbh]
\includegraphics[width=1.0\linewidth]{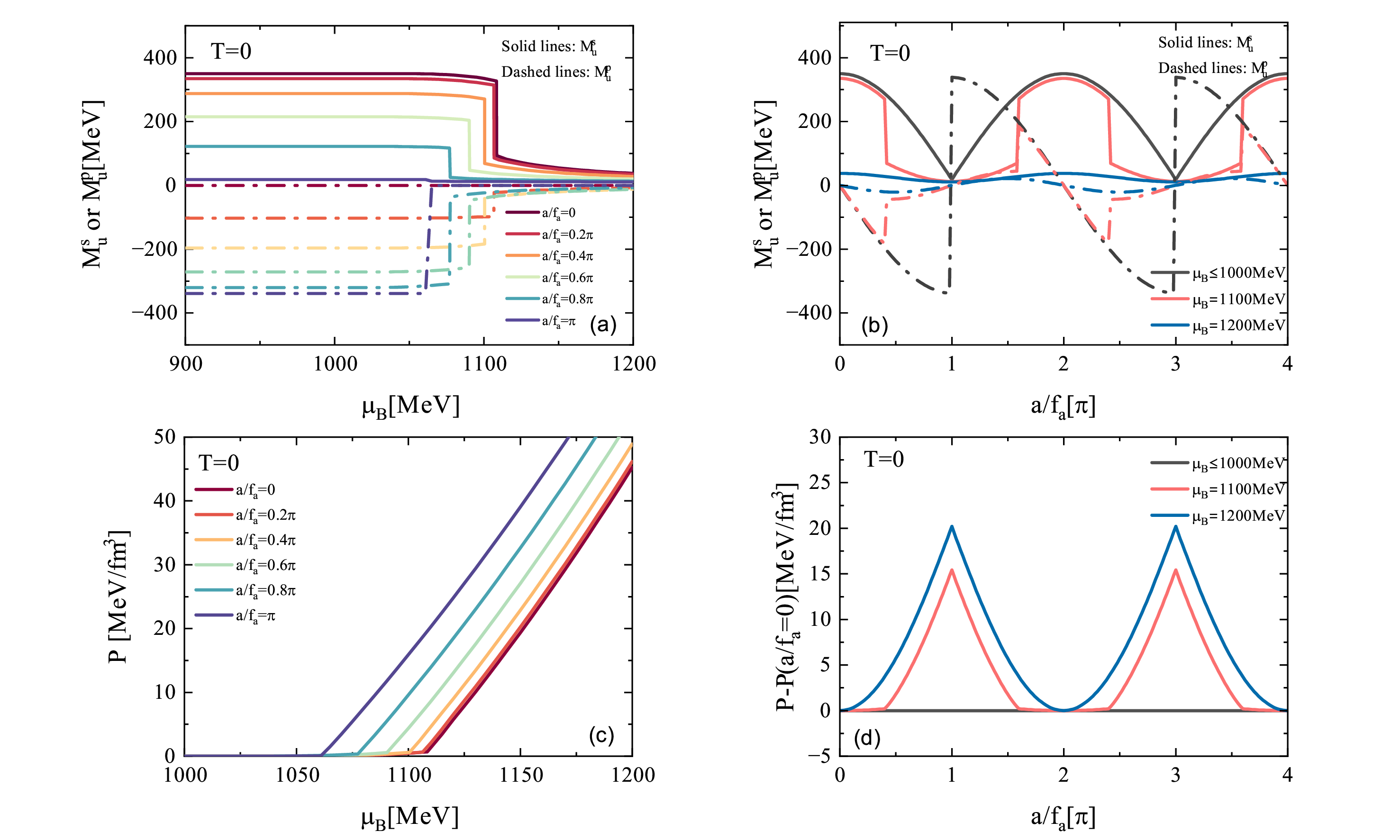}
\centering
\caption{ Axion effects on quark matter at zero temperature from NJL model with $G_V=0$, including (a) the scalar and pseudoscalar contributions of light quark constituent mass, $M_u^s$ and $M_u^p$, as functions of the baryon chemical potential for varying values of the scaled axion field $a/f_a$, (b) $M_u^s$ and $M_u^p$ as functions of the scaled axion field $a/f_a$ for several baryon chemical potentials, (c) pressure as a function of the baryon chemical potential for different values of the scaled axion field $a/f_a$, and (d) normalized pressure as a function of the scaled axion field $a/f_a$ for different baryon chemical potentials.}\label{fig1}
\centering
\end{figure*}

In this work, the hadron-quark mixed phase is described by Gibbs construction~\cite{Gle92,Gle01}. Including $\beta$-equilibrium, baryon number conservation and charge neutrality conditions, the dense matter enters the mixed phase, in which the nuclear and quark matter need to satisfy following equilibrium conditions:
\begin{eqnarray}
 P^H &=& P^Q,  \  \mu_i=\mu_Bb_i-\mu_cq_i, \\
\rho_B&=&(1-Y)(\rho_n+\rho_p)+\frac{Y}{3}(\rho_u+\rho_d+\rho_s),\\
0&=&(1-Y)\rho_p+\frac{Y}{3}(2\rho_u-\rho_d-\rho_s)-\rho_e-\rho_\mu,
\end{eqnarray}
where the labels $H$ and $Q$ represent the hadronic and quark phases, respectively, $\mu_B$ and $\mu_c$ are the baryon and charge chemical potential, as well as  $Y$ is the baryon number fraction of the quark phase. The core-crust transition density as well as the crust EOS in hybrid stars are also treated properly according to Refs.~\cite{Jxu09C,Jxu09J}. Using the whole EOS from hadronic to quark phase, the mass-radius relation of hybrid stars can be obtained by solving the Tolman-Oppenheimer-Volkoff (TOV) equation~\cite{Opp39}.

\begin{figure*}[tbh]
\includegraphics[width=1.0\linewidth]{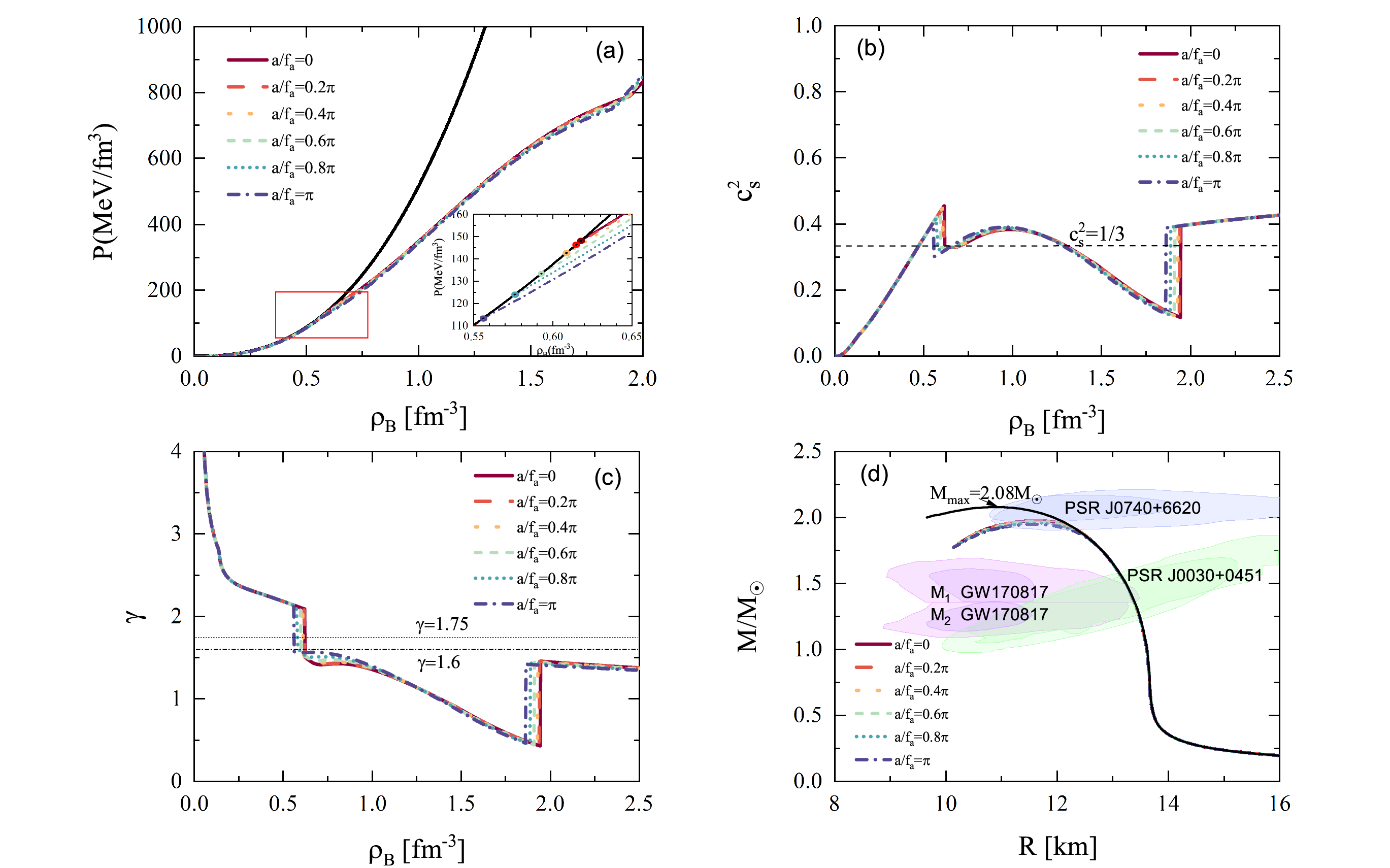}
\centering
\caption{Axion effects on hybrid star (matter) with the hadron-quark phase transition based on the NJL model for quark matter with $G_V=0.5 G_S$, and varying the scaled axion field $a/f_a$. The results include (a) pressure as a function of the baryon density,  (b) the squared of speed of sound $c_s^2$ as a function of the baryon density, (c) polytropic index $\gamma$ as a function of the baryon density, and (d) mass-radius relation. In (a), the color cycles indicate the onsets of the hadron-quark mixed phases, and the black solid line represents pure hadronic matter (PHM), which is also shown for comparison. In (d), constraints from multimessenger astronomy observations~\cite{Ril19,Mil19,Ril21,Mil21,Abb17} are shown by shaded regions, see text for details.}\label{fig2}
\centering
\end{figure*}

\begin{figure*}[tbh]
\includegraphics[scale=0.5]{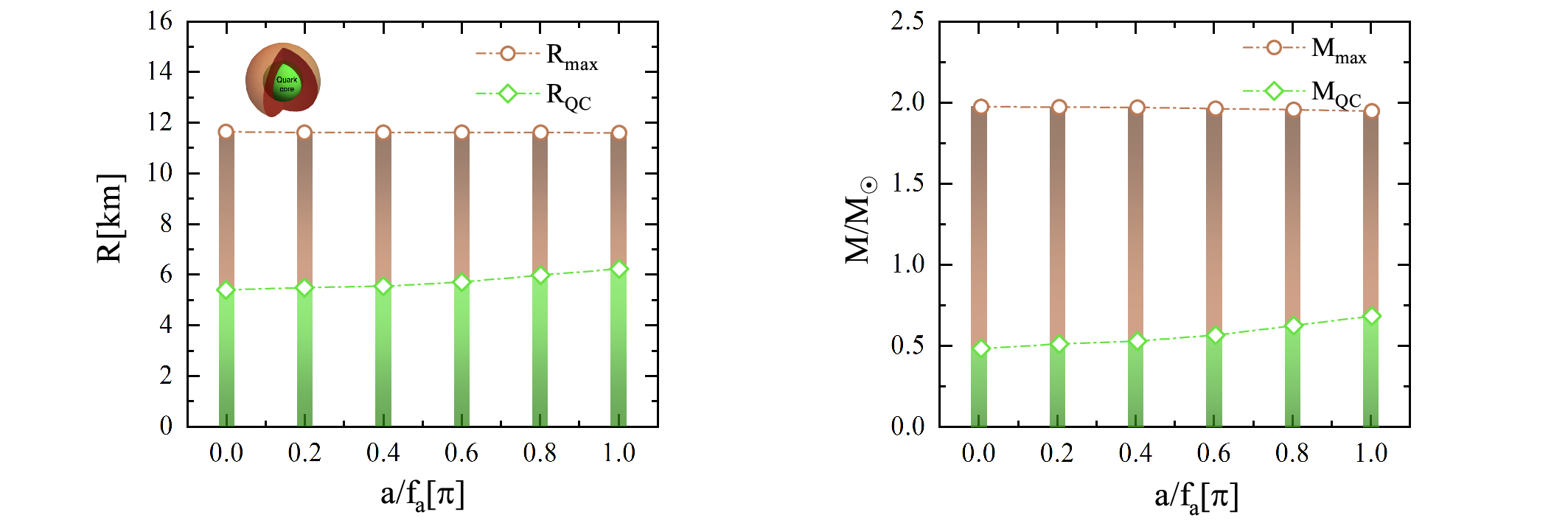}
\centering
\caption{ Radii and masses of quark-matter cores in the maxmum-mass hybrid stars for varying values of the scaled axion field $a/f_a$. The green bars represent the radii and masses of the quark-matter cores, while the chocolate-colored bars represent the radii and masses of the remaining parts. }\label{fig3}
\centering
\end{figure*}

\textit{Results and discussions.} We first display in Fig.~\ref{fig1} the axion effects on properties of quark matter at zero temperature from the NJL model with $G_V=0$. In the present computation, we assume isospin symmetry and set $\mu_u=\mu_d=\mu_s=1/3\mu_B$, thus ensuring that the properties of up quarks and down quarks among the light quarks are consistent. In Fig.~\ref{fig1}(a), we show the baryon chemical potential dependence of contributions to the constituent mass of up quark from the scalar and pseudoscalar condensates for varying values of the scaled axion field $a/f_a$. It can be clearly seen that within the range of $0$ to $\pi$ for $a/f_a$, the scalar mass $M_u^s$ decreases as $a/f_a$ increases, while the absolute value of the pseudoscalar mass $M_u^p$ increases with $a/f_a$. This forms a complementary relationship, resulting in the total constituent mass $M_u =\sqrt{{M_u^s}^2+{M_u^p}^2}$ remaining approximately constant for different values of $a/f_a$. Importantly, we can see that the first-order chiral phase transition is sensitive to the scaled axion field. Increasing $a/f_a$ decreases the baryon chemical potential of the first-order phase transition. In Fig.~\ref{fig1}(b), we show the variations of $M_u^s$ and $M_u^p$ with respect to the scaled axion field $a/f_a$ for several baryon chemical potentials. We can clearly see that $M_u^s$ and $M_u^p$ vary with $a/f_a$ in a periodic manner, with a period of $2\pi$. For $\mu_B \leq 1000$MeV, spontaneous CP violation is clearly evident when $a/f_a = \pi + 2k\pi$, with $M_u^p$ exhibiting two degenerate solutions that differ only in sign. For $\mu_B = 1100$ MeV, we can see the discontinuous changes in the magnitudes of $M_u^s$ and $M_u^p$, which are the results of the first-order chiral phase transition. For $\mu_B = 1200$ MeV, the restoration of chiral symmetry and CP symmetry leads to small and continuous changes in the magnitudes of $M_u^s$ and $M_u^p$. In Figs.~\ref{fig1} (c) and (d), we also show the axion effect on the pressure of quark matter. It can be seen that as the scaled axion field $a/f_a$ increases, the first-order phase transition in quark matter moves to the lower baryon chemical potential, resulting in the earlier appearance of free quarks, and consequently the equation of state of quark matter becomes stiffer. The normalized pressure given by $P-P(a/f_a=0)$ has a minimum value at $a/f_a=0$. After restoration of chiral symmetry and CP symmetry, the normalized pressure has a peak at $a/f_a=\pi+2k\pi$.

Next we discuss the axion effects on hybrid star (matter) with the hadron-quark phase transition based on the NJL model for quark matter with the different scaled axion field $a/f_a$. It is known that the position of the critical point for the chiral phase transition is sensitive to $G_V$~\cite{Asa89,Fuk08}, which was later constrained within $0.5 G_S \le G_V \le 1.1G_S$ from the relative $v_2$ splitting between protons and antiprotons as well as between $K^+$ and $K^-$ in relativistic heavy-ion collisions~\cite{Jxu14}. Therefore, we fix the coupling constant $ G_V = 0.5 G_S $, which can allow the maximum mass of the hybrid star to approach $2M_{\odot}$. We show in Fig.~\ref{fig2} (a) that the EOSs of hybrid star matter with the hadron-quark phase transition for the different scaled axion field $a/f_a$. The color cycles in the inset panel denote the onsets of the hadron-quark mixed phases. With increasing $a/f_a$, the EOS of pure quark phase in hybrid star becomes stiffer, which is consistent with that observed in Fig.~\ref{fig1} (c), whereas the shift of the mixed phase onset to lower densities actually softens the EOS of the mixed phase matter due to the earlier appearance of additional degrees of freedom. Using the EOS with the hadron-quark phase transition, we can determine the other properties of hybrid star matter. In Figs.~\ref{fig2} (b) and (c), we show that the squared speed of sound $c_s^2$ and the polytropic index $\gamma$ as functions of the baryon density in hybrid star by varying the scaled axion field $a/f_a$. The speed of sound and polytropic index are respectively defined as $c_s^2 \equiv \partial{P}/\partial{\varepsilon}$ and $\gamma\equiv d(\textrm{ln}P)/d(\textrm{ln}\varepsilon)$, which are considered to be good approximate criteria for the evidences of SQM in NS~\cite{Ann20,Han23,Liu232,Yan24}. As shown in Figs.~\ref{fig2} (b) and (c), we can observe a step-like decrease in both the speed of sound and the polytropic index during the hadron-quark transition, where the appearance of quarks softens the EOS. Subsequently, these two quantities exhibit another step-like increase as the nucleon and lepton degrees of freedom diminish in the high-density quark phase. It can also be seen that the speed of sound and the polytropic index in the pure quark phase at high densities exceed the conformal limits of $c_s^2 = 1/3$ and $\gamma = 1$, due to the strong repulsive vector interaction in quark matter. Our previous work~\cite{Liu231} indicated that if $G_V = 0$, both $c_s^2$ and $\gamma$ would converge to the conformal limit. Our results also agree with the approximate rules that $\gamma \leq 1.75$ ~\cite{Ann20} or $\gamma \leq 1.6$ and $c_s^2 \leq 0.7$~\cite{Han23} can be used as criteria for distinguishing hadronic matter from quark matter. The scaled axion field has slight impact on $c_s^2$ and $\gamma$ in mixed and pure quark phase, which indicates that after the appearance of quark matter, the effect of axions on the equation of state of hybrid star matter is mostly not dependent on density or energy density.

The mass-radius relation of the hybrid star with the scaled axion field is shown in Fig.~\ref{fig2} (d). The constraints from the bayesian analyses of the observational data from the pulsars PSR J0030+0451~\cite{Ril19,Mil19} and PSR J0740+6620~\cite{Ril21,Mil21}, and from the analyses of the gravitational wave signal from the NS merger GW170817~\cite{Abb17} are shown for comparison. The results indicate that both the observed maximum mass and the corresponding radius of hybrid stars slightly decrease with increasing $a/f_a$. This is due to the maximum mass of hybrid stars primarily constraining the EOS of mixed phase at densities in the range $2\rho_0 \sim 5\rho_0$. We also note that masses and radii of hybrid stars in all parameter sets are mostly consistent with the constraints from the pulsars PSR J0030+ 0451 and PSR J0740+6620, as well as the NS merger GW170817. The hybrid star matter consists of charge-neutral matter in $\beta$-equilibrium that has a hadron-quark phase transition from hadronic to quark matter. To better understand the axion effects on quark matter in hybrid stars, we show in Fig.~\ref{fig3} the radii and masses of quark-matter cores in the maxmum-mass hybrid stars for the different scaled axion field $a/f_a$. In this work, we consider the quark-matter core including the pure quark phase and the mixed phase. Using the TOV equations, the pressure is integrated from the central baryon density to the onset of the mixed phase to determine the size of quark-matter core. As shown in Fig.~\ref{fig3}, although the axion has a minimal impact on the mass and radius of the maximum-mass hybrid star, and even slightly reduces these values as $a/f_a $ increases, it significantly increases the radius and mass of the quark-matter core. This is due to the axion effect substantially lowering the density of onset of the mixed phase, while the central density of the maximum-mass hybrid star remains almost constant for different values of $a/f_a $. For the case $a/f_a = \pi $, the mass of the quark-matter core in a hybrid star with nearly $2M_{\odot}$ can reach up to $0.7M_{\odot}$, and the radius of core $R_{QC}=6.2$ km exceeds half of the star's total radius. Therefore, our results indicate that with axion effects, a sizable quark-matter core can appear in $2M_{\odot}$ massive neutron stars.

\textit{Summary and outlook.} In this work, we investigate the axion effects on quark matter and quark-matter core in massive hybrid stars based on the three-flavor NJL model. Results show that as scaled axion field $a/f_a$ increases within $0$ to $\pi$, the scalar mass component $M_u^s$ decreases while the pseudoscalar $M_u^p$ increases. The baryon chemical potential of the first-order chiral phase transition decreases with increasing $a/f_a$, causing earlier free quark appearance and stiffening the EOS for quark matter. The effect of axion on hybrid star matter is contrary to expectations. Axion field shifts the mixed phase onset to lower densities but slightly softens the EOS for the mixed phase matter, which marginally reduces hybrid star maximum mass and radius but significantly enlarges the quark-matter core's mass and radius. Consequently, our results suggest that axion effects enable the presence of a substantial quark-matter core in $2M_{\odot}$neutron stars. Considering axions as cold dark-matter candidates, our findings complement research into the impacts of beyond-Standard-Model particles on NS structure and binary mergers. Furthermore, we also observe a sudden steplike decrease in speed of sound and polytropic index at the hadron-quark phase transition, which agrees with the criteria ($\gamma \leq 1.75$~\cite{Ann20} or $\gamma \leq 1.6$ and $c_s^2 \leq 0.7$~\cite{Han23}) for distinguishing hadronic from quark matter. Our previous work~\cite{Liu24} indicates that $c_s^2$ and $\gamma$ can also be used to search for the QCD critical point in heavy-ion collisions, prompting further investigation into axion effects on the QCD phase diagram and critical point.

\textit{Acknowledgment.} This work is supported by the National Natural Science Foundation of China under Grants No. 12205158 and No. 11975132, and the Shandong Provincial Natural Science Foundation, China Grants No. ZR2021QA037, and No. ZR2022JQ04.

\end{document}